\documentclass[10pt,letterpaper,twocolumn]{article} 

\usepackage{ulem}
\usepackage{ol2} 
\usepackage[draft]{hyperref} 
\usepackage{amsmath}
\usepackage{amssymb}
\usepackage{epsfig}
\newcommand{\nm}{ \,{\text{nm}}}
\newcommand{\f}{ f }
\newcommand{\mum}{ \,\mu{\text{m}}}

\newcommand{\etaPL}{\eta_{\text{PL}}}
\newcommand{\etaPLm}[1]{\eta_{\text{PL},#1}}

\newcommand{\tslab}{t_{\text{ch}}}
\newcommand{\Wslab}{W_{\text{ch}}}

\newcommand{\Lc}{L_{\text{c}}}
\newcommand{\hEx}[2]{\text{hE}^{x,#1}_{1#2}}
\newcommand{\hEy}[2]{\text{hE}^{y,#1}_{1#2}}
\newcommand{\nre}{\text{Re}\left\{ n_{\text{eff}} \right\} }
\newcommand{\nim}{\text{Im}\left\{ n_{\text{eff}} \right\} }

\begin{document}

\twocolumn[

\title{Fiber-coupled semiconductor waveguides as an efficient optical interface to a single quantum dipole}


\author{Marcelo Davan\c co$^{* 1,2}$ and Kartik Srinivasan$^1$}
\address{$^1$Center for Nanoscale Science and Technology,
National Institute of Standards and Technology, Gaithersburg, MD
$^2$Maryland NanoCenter, University of Maryland, College Park, MD,
20742\\$^*$Corresponding author: mdavanco@nist.gov}

\begin{abstract}
We theoretically investigate the interaction of a single quantum
dipole with the modes of a fiber-coupled semiconductor waveguide.
Through a combination of tight modal confinement and phase-matched
evanescent coupling, we predict that $\approx70~\%$ of the dipole's
emission can be collected into a single mode optical fiber. We
further show that the dipole strongly modifies resonant light
transmission through the system, with over an order of magnitude
change for an appropriate choice of fiber-waveguide coupler
geometry.
\end{abstract}


]



The interaction of a single quantum dipole with a strongly confined
optical field is a central paradigm in quantum optics
\cite{ref:Kimble2}.  The ability to collect a large fraction of the
dipole's emission or use it to modify an incident optical field lies
behind a number of proposed applications in areas such as classical
and quantum information processing
\cite{ref:Kimble2,vanEnk.pra.69.043813,domokos.pra.65.033832,ref:Shen,ref:Chang-Lukin}
and single emitter spectroscopy \cite{ref:Gerardot2}. Such
applications depend on the availability of efficient and accessible
dipole excitation and emission channels. For instance, a single atom
in free-space is exclusively excited by the dipole wave component of
an illuminating field \cite{vanEnk.pra.69.043813}, and perfect
reflection of an illuminating directional dipolar field is expected
\cite{ref:zumofen}. Alternately, a single atom inside a Fabry-Perot
cavity is strongly excited by, and radiates efficiently into,
externally-accessible cavity modes and profoundly modifies the
resonator transfer function \cite{ref:Kimble2}.  Here, we
theoretically investigate a system in which a single emitter
embedded in a fiber-coupled semiconductor channel waveguide is
optically accessed with high efficiency, potentially yielding
$>70~\%$ fluorescence collection into a single mode optical fiber.
When resonantly interrogated, the dipole modifies the system's
transmission level by over an order of magnitude ($\approx15$ dB).

\begin{figure}[htfb]
\centerline{\includegraphics[width=8.3cm,trim=0 11 0 0]{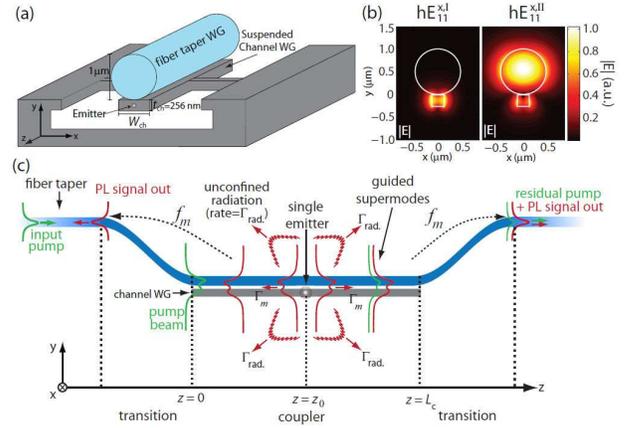}}
\caption{Fiber taper/channel WG directional coupler scheme. (a) 3D
schematic and coupler cross-section. (b) Hybrid $\hEx{\text{I}}{1}$
and $\hEx{\text{II}}{1}$ coupler supermodes for $\Wslab=190\nm$ and
$\lambda=1.3\mum$. (c) Non-resonant single dipole PL collection
configuration.} \label{fig-probing-scheme}
\end{figure}

Our system (Fig. \ref{fig-probing-scheme}) is an emitter embedded in
a suspended semiconductor channel waveguide (WG), evanescently
coupled to an optical fiber taper WG. The taper is a single mode
optical fiber whose diameter has been adiabatically and
symmetrically reduced to a wavelength-scale minimum, resulting in a
low-loss, double-ended device with standard fiber input and output.
Fiber and channel WGs form a directional coupler (cross-section
shown in Fig. \ref{fig-probing-scheme}(a)) of length $\Lc$, so power
may be transferred between the two guides. This system serves as an
efficient optical interface to a single dipole due to the
availability of a small number of WG modes with highly effective
coupling to the atomic transition (i.e., high
$\beta$-factors\cite{ref:rao,ref:LeCamp}), and access to such modes
via the fiber taper WG, which links on-chip nanophotonics and
off-chip fiber optics. As depicted in Fig.
\ref{fig-probing-scheme}(c), a signal launched into the fiber input,
adiabatically reduced in size along the fiber taper, excites
supermodes of the directional coupler. Guided supermodes illuminate
the WG-embedded dipole at position $z_0$ along the coupler. Upon
non-resonant excitation, the dipole emits coupler supermodes, at a
red-shifted wavelength, in the $\pm z$
directions\cite{ref:res_fluor_note}. Emitted supermodes are
converted into input and output fiber modes through the taper
transition regions, after which emission is detected. The individual
supermode contribution to the total photoluminescence (PL)
collection efficiency, $\etaPL$, is $\etaPLm{m}=f_m\cdot\Gamma_m /
\Gamma = f_m\cdot\gamma_m$, where $\Gamma_m$ is the supermode
emission rate, and $\Gamma$ the total emission
rate\cite{ref:Davanco}. The fraction $\gamma_m$ is supermode $m$'s
$\beta$-factor. Since emission in both $\pm z$ directions is equally
likely, $0\leq\gamma_m\leq0.5$. The fiber mode fraction, $f_m$, is
an overlap integral between the fundamental fiber mode and supermode
$m$~\cite{ref:Davanco,ref:Huang3}. Its quantum mechanical operator
analog is given here as Eq.~(\ref{eq-F-op}).

We study a geometry modeling a suspended GaAs channel with an
embedded self-assembled InAs quantum dot (modeled as a two-level
atom with electric dipole moment on the $xz$ plane) produced from
the material used in~\cite{ref:Srinivasan16}. The channel WG,
surrounded by air, has thickness $\tslab=256\nm$, width $\Wslab$,
and refractive index $n$=3.406 at a wavelength $\lambda=1.3\mum$.
The adjacent fiber has a $500\nm$ radius and $n$=1.45. For our
parameter range, the directional coupler region supports a set of
propagating supermodes named hE$^{x,y}_{11}$, hybrids of the
E$^{x,y}_{11}$ rectangular dielectric channel WG
modes~\cite{ref:rect_WG_modes} and fundamental fiber mode.
Supermodes hE$^{x}_{11}$ and hE$^{y}_{11}$ are excited by the $x$-
and $z$-electric dipole moment components, respectively.
Following~\cite{ref:Davanco}, where fiber-based collection of
emitters in membranes was studied, supermode field profiles
(calculated with the finite element method) were used to find
$\Gamma_m$ and $f_m$, while the finite-difference time-domain (FDTD)
method was used to calculate the total spontaneous emission rate
$\Gamma$.  These quantities allowed us to determine the total
fiber-collected PL efficiency ($\etaPL$) and individual supermode
contributions $\etaPLm{m}$. In addition, FDTD was used to obtain
$\etaPL$ without use of supermodes.

\begin{figure}[htfb]
\centerline{\includegraphics[width=8.3cm,trim = 0 20 0 0]{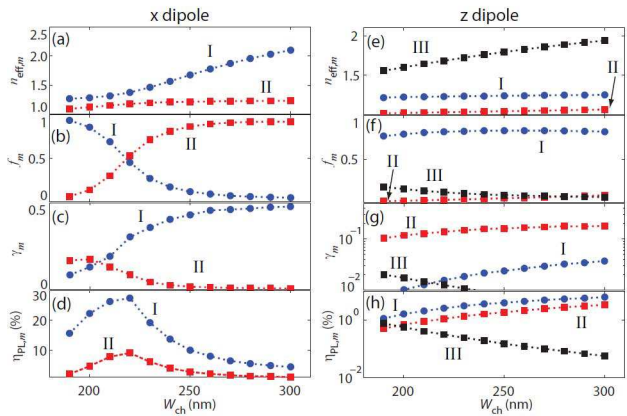}}
\caption{Effective index $n_{\text{eff},m}$, fiber mode fraction
$\f_{m}$, $\beta$-factor $\gamma_{m}$, and PL collection
contribution $\etaPLm{m}$ as functions of channel WG width $\Wslab$
for (a)-(d) $\hEx{m}{1}$ ($m$ = I or II) (e)-(h) $\hEy{m}{1}$
supermodes ($m$ = I, II, or III).}\label{fig-channel-gamma}
\end{figure}

Varying the channel width $\Wslab$ between $190\nm$ and $350\nm$
allows for significant modification of the supermode effective index
$n_{\text{eff}}$ . The real part of $n_{\text{eff}}$ for the
$\text{hE}^{\text{x}}_{11}$ doublet available in this range, labeled
$\text{I}$ and $\text{II}$, is shown in Fig.
\ref{fig-channel-gamma}(a). Both supermodes are guided, with
$\nim\approx10^{-11}$. Field profiles for $\Wslab=190\nm$ are shown
in Fig. \ref{fig-probing-scheme}(b). Phase-matching between the
fiber and E$^x_{11}$ modes is apparent near
$W_{\text{channel}}=220\nm$, where $f_{\text{I,II}}$ in Fig.
\ref{fig-channel-gamma}(b) are equal. As $\Wslab$ increases,
supermode $\hEx{\text{I}}{1}$ concentrates in the channel, resulting
in reduced $f_{\text{I}}$ and increased $\gamma_{\text{I}}$ (Fig.
\ref{fig-channel-gamma}(c)). Note, for $\Wslab\gtrapprox240\nm$,
$\gamma_{\text{I}}$ approaches the upper limit of $0.5$. For
$z$-oriented dipoles, two guided ($\nim\approx10^{-11}$)
$\text{hE}^{\text{y}}_{11}$ supermodes are available, labeled I and
III, with $\nre$ and $f_{\text{I,III}}$ plotted in Fig.
\ref{fig-channel-gamma}(e)-(f). A third supermode, (leaky,
$\nim\gtrapprox10^{-7}$), $\hEy{\text{II}}{1}$, has the highest
emission rate, though small $f_{\text{II}}$. Since the $y$-electric
field component is dominant, $\gamma_{m}$ for $z$-dipoles (Fig.
\ref{fig-channel-gamma}(e)) is small compared to the $x$-dipole
case. The highest contribution to $\etaPL$ is from the
$\hEx{\text{I}}{1}$ supermode, with
$\gamma_\text{I}\lessapprox0.04$.

\begin{figure}[htfb]
\centerline{\includegraphics[width=8.3cm,trim = 0 20 0 0]{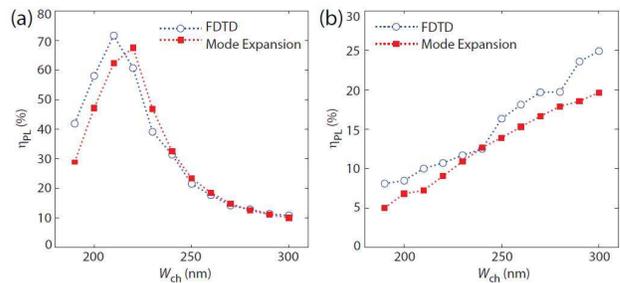}}
\caption{Maximum spontaneous emission collection efficiencies for
(a) $x$- and (b) $z$-polarized dipole moments,  as function of
$\Wslab$, calculated with FDTD and supermode expansion.}
\label{fig-channel-etaPL}
\end{figure}

Figure \ref{fig-channel-etaPL} shows total collection efficiency
$\etaPL$ for $x$- and $z$-oriented dipoles (including emission in
both $\pm z$ direcions, which is experimentally realizable),
obtained with FDTD and the supermode expansion method
of~\cite{ref:Davanco}. Since in each case multiple supermodes with
differing propagation constants are excited, the collection
efficiency oscillates along $z$, evidence of the power exchange
between channel WG and fiber. Collection maxima for $1\mum<z<5\mum$
are plotted for each $\Wslab$. The collection efficiency for
$x$-dipoles is maximized, nearing $70~\%$, for
$\Wslab\approx220\nm$.  Near the optimal point, most
($\gtrapprox73~\%$) of the emitted power is coupled into $\pm z$
propagating supermodes; fiber and slab are phase-matched, with equal
fiber fractions of $50~\%$, so the $\etaPLm{\text{I,II}}$ collection
contributions are maximized. For $z$-dipoles, a more modest $\etaPL$
is achieved, due to lower $\gamma_{m}$ and $f_{m}$ (Figs.
\ref{fig-channel-gamma}(f) and (g)). For $\Wslab=300\nm$, $\etaPL$
reaches $\approx25~\%$ , however the total rate $\Gamma$ is only
$\approx40~\%$ of that in the $x$-dipole case.

\begin{figure}[htfb]
\centerline{\includegraphics[width=8.3cm,trim = 0 20 0
20]{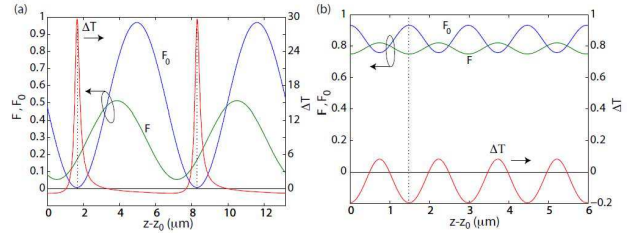}} \caption{Normalized, off- and on-resonance
transmission ($F_0$ and $F$) and contrast $\Delta T=(F-F_0)/F_0$ as
functions of separation from a single dipole at $z_{0}$, for (a)
$\Wslab=220\nm$ and (b) $\Wslab=300\nm$.} \label{fig-flux}
\end{figure}

We next show that the transmission through this directional coupler
can be significantly affected by the presence of the embedded
dipole.  Dipole-mediated control of light transmission has been
considered in studies involving free-space
\cite{vanEnk.pra.69.043813,ref:Gerardot2,ref:zumofen} and guided
mode excitation
\cite{domokos.pra.65.033832,ref:Shen,ref:Chang-Lukin}, and also in
cavity QED (e.g., \cite{ref:Kimble2,ref:Srinivasan16}). We start by
describing spontaneous emission as an electric dipole-type
interaction between the emitter and a vacuum field reservoir, given
in terms of the coupler supermodes \cite{ref:Davanco,
LeKien.pra.72.032509, vanEnk.pra.69.043813}. With the input-output
formalism of~\cite{vanEnk.pra.69.043813}, under the Markoff
approximation, we obtain the following steady-state,
positive-frequency, output multimode field operator for $z>z_0$
(i.e., past the dipole location):
\begin{eqnarray}
\mathbf{E}^{(+)}(z,t) = i\sqrt{2\pi}\sum_m
\sqrt{\frac{\hbar\omega}{4\pi
S_m}}\mathbf{e}_me^{-i(\omega t-\beta_mz)}\times \nonumber \\
\times\left[
\hat{a}^m_{in}(t-n_mz/c)+\sqrt{\Gamma_m}^*\sigma_-(t-n_mz/c)
\right]. \label{eq-e-field-op}
\end{eqnarray}
Here, $\sigma_-$ is the atomic lowering operator, $\hat{a}^m_{in}$
is supermode $m$'s input field annihilation operator, $\mathbf{e}_m$
is the electric field distribution, $\beta_{m}$ the propagation
constant, $n_m$ the phase index, and $S_m=\text{Re} \{ \int_S\,dS(
\mathbf{e}_m\times\mathbf{h}^*_m )\cdot\mathbf{z} \}$, with $S$ the
$xy$ plane. The expression in brackets is a well-known result of the
input-output formalism, with explicit input (or "free") field and
radiated ("source") field contributions
~\cite{vanEnk.pra.69.043813}. Next, we assume the percentages of
incident fiber mode power transferred to coupler supermodes at $z=0$
are given by the fiber-mode fractions $f_m$, and that the power
coupled into the output fiber at $z=\Lc$ is approximated by an
overlap integral between the field at this position and the fiber
mode (Eq. (2) in~\cite{ref:Davanco}). This expression is translated
into the fiber power operator
\begin{eqnarray}
\hat{F}=\left\{
\int_SdS(\mathbf{E}^{(-)}\times\mathbf{h}_f)\cdot\mathbf{z}
\int_SdS(\mathbf{e}^*_f\times\mathbf{H}^{(+)})\cdot\mathbf{z}+ \right. \nonumber \\
\left.\int_S\,dS(\mathbf{H}^{(-)}\times\mathbf{e}^*_f)\cdot\mathbf{z}
\int_SdS(\mathbf{h}^*_f\times\mathbf{E}^{(+)})\cdot\mathbf{z}
\right\}S^{-1}_f, \label{eq-F-op}
\end{eqnarray}
where $\mathbf{e}_f$ and $\mathbf{h}_f$ are the fiber mode electric
and magnetic field distributions, and $S_f=\text{Re} \{ \int_S\,dS(
\mathbf{e}_f\times\mathbf{h}^*_f )\cdot\mathbf{z} \}$. Photon flux
and higher order correlation functions at the output fiber may be
obtained with $\hat{F}$. Using Eq.~(\ref{eq-e-field-op}) into
Eq.~(\ref{eq-F-op}) and assuming a coherent state illumination
source, an expression for the output fiber photon flux expectation
value is obtained in the low-excitation limit (far below
saturation), and normalized to the input photon flux $F_{\text{in}}$
to produce the transmission level $F =
\langle\hat{F}\rangle/F_{\text{in}}$. The resulting $F$ expression
consists of a sum of terms proportional to $f_m
f^*_{m'}e^{i(\beta_m-\beta_{m'})(z-z_0)}$, and is used to calculate
the transmission contrast through the fiber, defined as ${\Delta}T =
(F-F_0)/F_0$, where $F$ and $F_0$ are the transmission levels on and
off resonance with an $x$-polarized dipole. The transmission
contrast is significant over a bandwidth of the order of the
transition linewidth (the Purcell enhancement is small in these
structures), which is much smaller than the coupler transmission
bandwidth. In Fig. \ref{fig-flux}, we plot $F_0$, $F$, and ${\Delta
}T$ for a coupler with $\Wslab=220\nm$ (phase-matched channel and
fiber WGs). As expected for a directional coupler, $F_0$ oscillates
along $z$ between close to zero and close to unity, with beat length
$L_\pi=\pi/(\beta_\text{I}-\beta_\text{II})\approx3.3\mum$. The
coupler 3 dB transmission bandwidth is $>100 \nm$ for
$\Lc\lessapprox5\mum$. For a dipole located at $z_0=L_\pi/2$, $F$
can be significantly enhanced or suppressed relative to $F_{0}$,
depending on $z$: at $z-z_0\approx1.65\mum$ ($L_\pi/2$),
$F\approx20~\%$ is nearly 30 times larger than $F_{0}<1~\%$; at
$z-z_0\approx5.0$ $\mu$m, $F\approx40~\%$ is 2.4 times smaller than
$F_{0}\approx96~\%$. A judicious choice of coupler length $\Lc$ thus
produces structures in which a single dipole strongly affects
transmission. This could enable, e.g., measurements of emitter
spectral diffusion, or, with AC or DC Stark effect emitter frequency
control, dipole-controlled light modulation. We note that phase
matching is crucial in such devices, as ${\Delta}T$ is limited by
incomplete power transfer in phase-mismatched fiber and channel WGs.
Figure \ref{fig-flux}(b) shows more modest results for
$\Wslab=300\nm$, due to phase mismatch ($\Delta T\approx-20~\%$ may
still be achieved).

If a single coupler supermode is accessed by the fiber, i.e.,
$f_m=0$ for all but one supermode, we find $F =
f_m^2\left[1-4\gamma_m(1-\gamma_m)\right]$, which illustrates the
essential role of $\gamma_m$ in extinction measurements
\cite{vanEnk.pra.69.043813}. Perfect extinction is predicted for
$\gamma_m=0.5$, or exclusive $m$-supermode emission. As emission is
in both directions, this is equivalent to perfect
reflection\cite{ref:zumofen}. In Fig.~\ref{fig-channel-gamma}(c), it
is apparent that extinction near $100~\%$ may be achieved for
$\Wslab >250\nm$, provided only supermode $\hEx{I}{1}$ is
accessible. This situation can be approximated with WG mode
conversion structures (e.g., lateral or vertical tapers) that favor
coupling between the fiber mode and specific coupler
supermodes~\cite{ref:Huang3}. For example, for $\Wslab=300\nm$, a
modest 80:20 coupling ratio to the type I and II supermodes (i.e.,
$f_\text{I}=0.8$, $f_\text{II}=0.2$, $f_{m\neq \text{I,II}}=0$)
would lead to $>88~\%$ extinction, independent of dipole position
and coupler length.

In summary, we have investigated a hybrid waveguide structure in
which strong dipole excitation is combined with efficient optical
access through an evanescently-coupled optical fiber-based
waveguide. These devices may have application in areas such as
quantum information processing and single emitter spectroscopy.

This work was partly supported by the NIST-CNST/UMD-NanoCenter
Cooperative Agreement.


\begin{thebibliography}{10}
\bibitem{ref:res_fluor_note}
Resonant fluorescence can also be studied, for example, through
emission into
  the backward traveling WG mode.
\bibitem{ref:rect_WG_modes}
Far from cutoff, channel WG E${^{x,y}_{pq}}$ modes are predominantly
$x$- or
  $y$-polarized, with $p$ and $q$ maxima in the $x$- and $y$-directions
  respectively.
\end{thebibliography}


\end{document}